\documentclass[twocolumn,showpacs,preprintnumbers,amsmath,amssymb]{revtex4}
\usepackage{graphicx}
\usepackage{dcolumn}
\usepackage{bm}

\begin{document}

\title{ Renormalized atomic  interaction  and quadrupole \\excitations  
 of cold Fermi gas near Feshbach resonance}
 \date{\today}

\author{Lai-Wa\ Siu, T.\ T.\ S.\ Kuo, B.\ A.\ Gelman}
\affiliation{Department of Physics and Astronomy, SUNY, 
Stony Brook, NY 11794-3800, USA}

\begin{abstract}
We present a model space particle-hole Green's function calculation for
the   quadrupole excitations 
of cold Fermi gas near Feshbach resonance using  a simple model where 
atoms are confined in a harmonic oscillator potential. Both the
Tamm-Dancoff and random phase approximations are employed.
By summing up exactly the ladder diagrams between a pair of interacting atoms
to all orders, we  obtain a renormalized atomic  interaction
which has well defined and  identical limits as the scattering length tends to
$\pm \infty$. The experimentally observed abrupt rise in the excitation 
spectrum and its associated large decay width
are satisfactorily reproduced by our calculation.


\end{abstract}

\pacs{pacs} \maketitle

{\bf   Introduction}: The Feshbach resonance was originally used in nuclear 
physics \cite{fesh62}, to describe a resonance in neutron scattering 
due to the formation of a quasi-bound state between the scatterer 
and the incident particle. 
During the past several years, this resonance was experimentally
realized in ultracold alkali gases 
\cite{joch02,regal03}. It has many interesting features
and has opened a new era to the study of ultracold gases. It serves
as an experimental control mechanism to tune 
the scattering length over a wide range,
essentially from positive to negative infinity. Through such
process, the inter-atomic effective interaction could also be
experimentally controlled.  As a result, ultracold 
alkali gases have provided us a unique many-body system with tunable
interactions, and they
have been the focus of many theoretical studies 
(See, e.g. \cite{pethick,duin04}). 
As the magnetic field is tuned across the Feshbach
resonance, ultracold Fermi gases have been observed to undergo the well known
 BCS-BEC crossover \cite{grein03,zweir04,barten04}.

Collective oscillations of trapped cold Fermi gas 
have  been recently measured \cite{barten04b,kinast04}, with important
and very interesting findings for the collisional-, collisionless- 
and particularly
the transitional-regime in between. In the collisional regime, 
the oscillation frequency is lower than the trap frequency and
it varies continuously across the Feshbach resonance where
the scattering length $a_s$ tends to $\pm \infty$.
In the collisionless regime, the frequency is essentially constant,
close to the trap frequency. In the transitional regime, however, an abrupt
rise of frequency together with a strong damping rate were observed.
To explain these findings has been a challenging task.
In the collisional regime, 
pioneering hydrodynamical studies 
 \cite{bertsch05,boris05}  have been carried out.

In this Letter, we would like to carry out a microscopic study  for 
the quadrupole excitations of trapped cold Fermi gas, aiming 
primarily at the transitional and collisionless regimes. 
 We shall use a particle-hole 
Green's function framework which has been commonly used in nuclear physics 
 \cite{speth91,wukuo84,blomq70}. When applying this method to nuclear
systems, nucleons are placed in an oscillator well and particle-hole
excitations are treated to all orders using either the Tamm-Dancoff
or random-phase approximations, commonly referred to as TDA and RPA
respectively.
 We shall use a similar approach for the above gas system,
 with its atoms  trapped in a harmonic
oscillator potential.

In the following section, we shall first  briefly describe the 
model space particle-hole
Green's function method \cite{wukuo84} which we shall employ.
 A specific feature of the atomic interaction for cold Fermi gases  is that
it is divergent at Feshbach resonance. 
\cite{stoof96,leggett01,pethick,heise02,kolck02}
This interaction is not suitable for being used directly in 
microscopic calculations; it needs to be
renormalized (or tamed) beforehand. We shall discuss that the well known
Brueckner reaction matrix method, which has been
widely and successfully used in nuclear matter 
\cite{jeuke75,kuo86,holt05} and finite nuclei \cite{hjorth95}, 
is very useful in this regard. We shall use this method to derive
a renormalized atomic interaction which is smooth and  well behaved
 at the Feshbach resonance. 
 Our results together with some calculational details will be reported 
and discussed in the last section. 
 A summary will also be presented.

{\bf Model space particle-hole Green's function method}:
In this section we briefly describe the model space particle-hole ($ph$) 
Green's function method \cite{wukuo84} for 
calculating the quadrupole excitations
of cold Fermi gas.  We consider
atoms as spin 1/2 fermions confined in a spherical harmonic oscillator
potential of oscillator spacing $\hbar \omega_{osc}$ and oscillator length
$a_{osc}=\sqrt{\hbar /m\omega_{osc}}$. In our present calculation, 
we assume the atoms remain normal throughout. As pointed out in
\cite{boris05}, in trapped fermi gas experiments, the superfluid
component always resides in the interior of the trap and had little 
contribution to the quadrupole excitation. We shall
use $\hbar \omega_{osc}$ and $a_{osc}$ respectively as 
the units for energy and length.

We treat the trapped gas as a closed shell system of  $N_F$ closed shells. 
For example, the $N_F=6$ system has 112 atoms
and in  its unperturbed ground state  the lowest 6 oscillator shells are 
completely filled.   The excitation energies, denoted by $E_n$, of the system
are calculated
 by solving the $ph$ RPA equation \cite{wukuo84,blomq70}
\begin{eqnarray}
    AX_n+BY_n &=&E_n X_n,  \nonumber \\
   -B^*X_n -A^*Y_n&=&E_n Y_n.
\end{eqnarray}
Here $A$ represents the sum of the unperturbed $ph$ energy gap
and the TDA vertex function, namely
\begin{equation}
A(ph,p'h')=(\epsilon_p+S(p)-\epsilon_h-S(h))\delta _{ph,p'h'}+\Sigma (ph,p'h'),
\end{equation}
where $S$ denotes the one-body vertex function and $\Sigma$ the two-body
$ph$ vertex function. They are composed of irreducible diagrams as illustrated
in Fig. 1. Here each dashed-line  represents the vertex of the effective
atomic interaction $V$. Note that the diagrams are time ordered
with retarded single particle propagators \cite{wukuo84}.
 $B$ denotes the ground-state correlation vertex, whose diagrammatic 
structure is also illustrated in Fig. 1.  All the diagrams
contained in the vertex function must be irreducible in the sense that
any intermediate state of  such diagrams must  be orthogonal
to the $ph$ model space \cite{wukuo84}. 

\begin{figure}[here]\label{f1}
\scalebox{0.4}{
\includegraphics{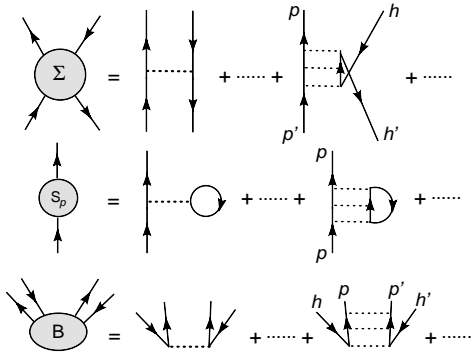}
}
\caption{Diagrams  for one- and two-body vertex functions.
The dotted line vertex represents a $V$ interaction. }
\end{figure}

The transition amplitudes  $X$ and $Y$ of Eq.(1) are 
$X_n(ph)=\langle \Psi _n |a_p^{\dagger}a_h |\Psi_0 \rangle$ and
$Y_n(ph)=\langle \Psi _n |a_h^{\dagger}a_p |\Psi_0 \rangle$,
where $\Psi_n$ and $\Psi_0$ represent  the $n$th excited- and ground-state
wave functions respectively. When the ground state correlation vertex
$B$ is suppressed, Eq.(1)  becomes the TDA equation.
 We consider that the atomic gas is trapped by
a harmonic oscillator potential. Thus  $\epsilon$ of Eq.(2) 
is the  harmonic oscillator single particle energy.

In carrying out the RPA/TDA calculations, a restricted model space
near the Fermi surface is employed.  We use a hole space
consisted of three major shells immediately below the Fermi surface,
 namely shells $N_F$ to $(N_F-2)$, and for particles
 one major shell immediately above. The external lines $p,h,p'$ and $h'$
of the vertex function diagrams of Fig. 1 are all confined in this space.
 That we use a restricted
model space requires a model-space renormalization, so that the effects
from the orbitals outside the model space are taken care of. The inclusion
of the higher order diagrams in the vertex functions is for this 
renormalization purpose. As an example, let us consider the third order 
diagrams of Fig. 1. They all have
a common ``ladder" structure, with repeated interactions between a pair
particle lines. The intermediate states of them  include all the particle
states outside  the model space, so that our $ph$ calculation
within a small model space have renormalization contributions from the
space outside. As discussed later, this
type of ``ladder" diagrams can be summed up to all orders, giving rise
to the reaction matrix interaction.

{\bf Separable reaction matrix interaction}:
In carrying out our 
microscopic particle-hole calculations,
 we must first
have an atomic interaction on which our many-body calculation
 is based. 
This interaction \cite{stoof96,leggett01,pethick,heise02,kolck02}  
has a specific form and is commonly given as 
\begin{equation}
V=4\pi \frac{\hbar^2}{m}a_s \delta(r),
\end{equation}  
where $m$ denotes the atomic mass, and $a_s$ is the s-wave scattering
length which can be tuned
across the Feshbach resonance, essentially from $-\infty$ to $+\infty$. 
The above is an effective interaction for one hyperfine channel, in the sense
that we consider  atoms as of only one hyperfine state with the other
hyperfine states having been integrated out by renormalization.
In our calculation we assume atoms interacting with s-wave interactions only.

The above interaction   is clearly  divergent  at the
Feshbach resonance (the 
unitary limit) where $a_s \rightarrow \pm \infty$, and 
there is  difficulty in using it in microscopic  calculations.
As described in section 2, to carry out the TDA/RPA calculations,
we need first calculate the vertex functions $S$, $\Sigma$ and $B$. As
illustrated in Fig. 1, these vertex functions are composed of irreducible
diagrams. 
Near the Feshbach resonance, the above atomic interaction is divergent
and consequently every diagram in the figure is divergent; calculation 
can not go on unless we use a different approach.
 This type of difficulty has been well known in nuclear matter theory
and has been overcome by the 
Brueckner reaction matrix theory 
\cite{jeuke75,kuo86,holt05}. There each vertex of the bare nucleon-nucleon
interaction $V_{NN}$ is nearly divergent because of its very strong repulsive
core. However, the Brueckner reaction matrix $G$ given by the  all order 
summation of the $V_{NN}$ interactions
between a pair of nucleons is finite and well behaved.

We apply a similar reaction matrix approach to our 
present atomic gas calculation. Let us consider the diagrams
for the one-body vertex function $S$ of Fig. 1. Clearly they have the 
structure of a geometric series and can be summed up to all orders.
Let us define a reaction matrix $R$ as
\begin{equation}
R=V+V\frac{Q}{e}V+V\frac{Q}{e}V\frac{Q}{e}V+...=V+V\frac{Q}{e}R,
\end{equation}
where $e$ stands for $(\omega-H_0+i0^+)$ and $Q$ the Pauli exclusion
operator. (Recall that we use time-ordered diagrams with
retarded single particle propagators.) $H_0$ is the unperturbed 
Hamiltonian for the pair of interacting atoms, and $\omega$ 
is the starting energy which we shall discuss later. 
The role of $Q$ is to ensure that the intermediate states of $R$
must be outside of the chosen model space.
In terms of $R$, the sum of all the ladder-type diagrams as shown by $S(p)$
of Fig.1 is now $\langle ph|R|ph \rangle$. Similarly the sum for $\Sigma$
is $\langle ph'|R|p'h \rangle$ and for $B$ is $\langle pp'|R|hh'\rangle$.
 We shall include only diagrams first order
in $R$ for the vertex function as listed in Fig.2.

\begin{figure}[here]\label{f2}
\scalebox{0.4}{
\includegraphics{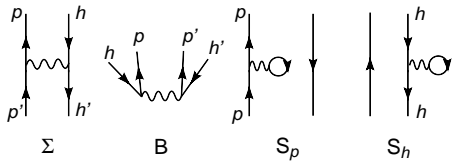}
}
\caption{First-order R-matrix (wavy-line vertex) diagrams  for one- and 
two-body vertex functions.}
\end{figure}
In calculating $R$, we  shall use an 
angle average approximation \cite{kuo86}
for the Pauli operator, namely $Q$ is approximated by $Q_{av}(q,k_F)$
where $q$ is the relative momentum  and $k_F$ is an average
Fermi momentum which can be estimated by using for example
a local density approximation.
In addition, we shall use plane-wave intermediate states and an
average starting energy $\omega$. 
Accurate methods for calculating the nuclear reaction matrix have been 
developed \cite{hjorth95}, and we plan to use them in a future work.
With these approximations, the above $R$ matrix becomes
\begin{eqnarray}
R(k,k',\omega)=V(k,k')~~~~~~~~~~~~~~~~ \nonumber \\
+\frac{2}{\pi}\int _0 ^{\infty}
V(k,q)\frac{Q_{av}(q,k_F)}{\omega-q^2+i0^+}R(q,k',\omega)q^2dq,
\end{eqnarray}
where $k,k',q$ are the relative momenta.

Let us now consider the above $R$ matrix for the scattering length
dependent potential of Eq.(3). 
Schematically, $R$ is given by $R=V/(1-VQ/e)$ and may have
 well defined limits 
as $V \rightarrow \pm \infty$. But how to obtain them accurately
is a difficult task in general, and it may not be feasible to obtain them 
reliably using numerical methods. We feel that $R$ has to be solved analytically
 in order to obtain the above  limits reliably. 
In this regard, we resort to the separable
potential approach which has been widely used in nuclear physics
 \cite{brownsepa,tabakin64,ernst73}; the reaction matrix with this
approach can be analytically solved. We shall use a  separable
potential which corresponds to a simulation of
the potential of Eq.(3) by a short range non-local Yukawa potential
of the form $a_s e^{-\mu r}e^{-\mu r'}/(\mu^2 rr')$,
 $\mu ^{-1}$ being the range of the interaction and $r$ and $r'$
the inter-atomic radial distance. We shall use large $\mu$ to simulate
short range potential.   In momentum space this potential becomes 
\begin{equation}
V(k,k')=a_s \alpha f(k)f(k');~f(q)=\frac{1}{\mu^2+q^2},~
\alpha=8\pi \frac{\hbar^2}{m}\frac{\mu ^3}{a_{osc}}.
\end{equation}

With this separable potential,
 the reaction matrix, which is complex, is also separable and is
obtained analytically as
\begin{equation}
R(k,k',\omega)=[\eta_{re} (\mu,\omega)+i\eta_{im} (\mu,\omega)]f(k)f(k'),
\end{equation}
with
\begin{eqnarray}
\eta_{re}(\mu,\omega)=\frac{\lambda(\mu,\omega)}{D(\mu,\omega)},
\nonumber \\
\eta_{im}(\mu,\omega)=\frac{-\lambda(\mu,\omega)^2 \sqrt{\omega}}
{D(\mu,\omega) (\mu^2+\omega)^2}, \nonumber \\
D(\mu,\omega)=1+\frac{\omega \lambda(\mu,\omega)^2 }
{ (\mu^2+\omega)^4},
\nonumber \\
\lambda(\mu,\omega)=\frac{a_s\alpha}{1-a_s\alpha I(\mu,\omega)}.
\end{eqnarray}
  Note that $\eta _{im}=0$ if $\omega<0$. In the above, $I$ is the integral
\begin{equation}
I(\mu,\omega)=\frac{2}{\pi}P\int_0 ^{\infty} \frac{Q_{av}(k,k_F)}{(\omega-k^2)
(\mu^2+k^2)^2}k^2dk,
\end{equation}
where $P$ denotes principal value integration. In a recent study
of fermionic atoms in optical lattices \cite{luu06},
 a Gaussian separable
interaction is employed and its strength has similar 
dependence on $a_s$ as the above $\lambda$.

It is readily seen that  $R$ has well defined limits as $a_s
\rightarrow \pm \infty$.  Only the factor $\lambda$ of Eq.(8) is dependent
on $a_s$, and at the above limits it becomes
\begin{equation}
\lambda _{\pm \infty}(\mu,\omega)=\frac{-1}{I(\mu,\omega)}.
\end{equation}
 Thus our $R$-matrix has a common
well defined limit at the
Feshbach resonance, being the same whether 
$a_s \rightarrow +\infty$ or $a_s \rightarrow -\infty$.

For dilute systems, $k_F$ should be small.
  We have calculated the above $R$ matrix for several values of $k_F$,
and found that the results for $R$ calculated with $k_F$ ranging 
from 0 to 1 are quite close to each other. For the case 
 of $k_F=0$ and $\omega >0$, the  integral of Eq.(9) becomes
\begin{equation}
I(\mu,\omega)=\frac{\omega-\mu^2}{2\mu(\omega+\mu^2)^2},
\end{equation}
which is only weakly energy dependent if $\omega$ is much smaller than
$\mu^2$. Thus our $R$-matrix interaction is nearly  energy 
independent for cold (small $\omega$)
dilute (small $k_F$) system  with a short-range (large $\mu$) interaction. 

\begin{figure}[here]\label{f4}
\scalebox{0.45}{
\includegraphics{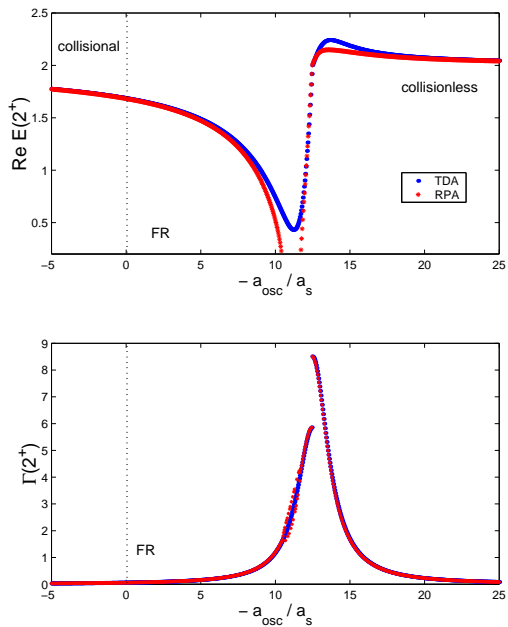}
}
\caption{Energy and decay width of the quadrupole excitation.}
\end{figure}


{\bf Results and discussion}:
 We now describe some details about our calculations for the 
energy and decay width of the quadrupole excitation. They are obtained
from the solutions of the RPA/TDA equation (1), using the vertex functions
given by the diagrams of Fig. 2. These diagrams are calculated from
the $R$-matrix interaction, using the methods detailed in \cite{hjorth95}.
Since $R$ is complex, Eq.(1) is now a complex equation and 
its eigenvalues $E_n$ are generally complex. We  write $E_n$ 
as $(ReE_n -i\Gamma _n)$, with the decay width given by $\Gamma _n$.
In Fig.3 we present  results for the lowest $2^+$ state of an $N_F$=8 
(240 atoms) model system, using 
parameters  $k_F$=0.3, $\omega$=1 and $\mu$=20. (We have also calculated
 $N_F$=6 and 7 systems, with results both being  highly similar to those
of $N_F=8$.) These parameters
should be suitable for cold dilute systems interacting with a short
range interaction. 
The above choice  has the advantage of 
making $R$-matrix weakly energy dependent as we have $\mu^2 >> \omega$.
As shown, our calculated energy at Feshbach resonance (FR) is about 1.7
$\hbar \omega_{osc}$
with nearly zero width, both being smooth functions there.

 The shapes of our calculated spectrum and decay width  are worth noting.
Starting from the left end of Fig. 3, $ReE$ exhibits a gradual  drop
followed by an abrupt rise and finally remains nearly constant. At both ends,
 $ReE$ approaches to 2. Most importantly,  the
decay width $\Gamma$ is peaked concurrently with the abrupt
rise, both happening right at $a_{\lambda}$. These specific features 
of our results  agree 
remarkably well with experiments \cite{barten04b,kinast04}.

The above abrupt rising feature is mainly due to the $R$-matrix interaction.
 The factor $\lambda$ of Eq.(8) has a pole
when $a_s$ equals to  $a_{\lambda}=-1/(\alpha I)$.  It is readily seen  that at
$a_{\lambda}$ the imaginary part of the $R$ matrix is peaked with a
Lorentzian form, while its real part vanishes. In addition, $R$ is positive
to the left of $(-a_{osc}/a_{\lambda})$, and negative on the other side. 
These properties of $R$ determine the  general shapes of our results.
For small $k_F$ and $\omega$,
 $a_{\lambda}$ is  close to $a_{osc}/4\pi$. In fact for $k_F$=0 
and $\omega=0$, $a_{\lambda}$ is exactly equal to $a_{osc}/4\pi$, 
independent of $\mu$. As shown
in the figure, the abrupt rising takes place very close to $4 \pi$.
 We note that
RPA and TDA give nearly identical results far from $a_{\lambda}$.
Near $a_{\lambda}$, $ReE$ for RPA tends to negative indicating the
system being unstable.

Last but not least,  
$\Gamma$ is discontinuous at $a_{\lambda}$ as shown. We have found that
the wave functions $X$ and $Y$ of Eq. (1) 
are also discontinuous at $a_{\lambda}$. 
 We plan to further study 
the nature of this ``transition" in a future work.

In summary, we have carried out a particle-hole shell model calculation
for the quadrupole excitations of trapped cold Fermi gas. An essential step
is the use of a reaction matrix renormalized interaction.
Using a separable potential approach, this interaction has been derived
analytically and it is  a smooth function at Feshbach resonance although
the atomic interaction before renormalization is divergent there.
Our results demonstrate that, in between the collisional and collisionless
regimes, trapped cold Fermi gas undergoes a transition with
 abrupt variations in both energy (frequency) and decay width 
(damping rate), in good qualitative agreement with experiments. 


{\bf Acknowledgement:}
We sincerely thank G.E. Brown, E. Shuryak and Tom Bergman
for many helpful discussions.
Partial support from the US Department of Energy under contract
DE-FG02-88ER40388 is gratefully acknowledged.



\end{document}